\documentclass[twocolumn,showkeys,aps,prb,showpacs]{revtex4-1}
\UseRawInputEncoding
\usepackage{graphicx}
\usepackage[CJKbookmarks,dvipdfm,colorlinks,linkcolor=red,citecolor=blue]{hyperref}

\begin{document}

\title{Sensitive  electronic correlation effects on electronic properties  in ferrovalley  material Janus  FeClF monolayer}

\author{San-Dong Guo$^{1}$, Jing-Xin Zhu$^{1}$,  Meng-Yuan Yin$^{1}$  and  Bang-Gui Liu$^{2,3}$}
\affiliation{$^1$School of Electronic Engineering, Xi'an University of Posts and Telecommunications, Xi'an 710121, China}
\affiliation{$^2$ Beijing National Laboratory for Condensed Matter Physics, Institute of Physics, Chinese Academy of Sciences, Beijing 100190, People's Republic of China}
\affiliation{$^3$School of Physical Sciences, University of Chinese Academy of Sciences, Beijing 100190, People's Republic of China}
\begin{abstract}
The electronic correlation may have essential influence on electronic structures in some materials with special structure and  localized orbital distribution. In this work,  taking Janus  monolayer FeClF as a concrete example,  the correlation effects
on its electronic structures are investigated by using  generalized gradient
approximation plus $U$ (GGA+$U$) approach. For perpendicular magnetic anisotropy (PMA),  the increasing  electron correlation effect can induce the ferrovalley (FV) to half-valley-metal (HVM)  to quantum anomalous Hall (QAH) to HVM to FV transitions. For QAH state, there are a unit Chern number and a chiral edge state connecting the conduction and valence bands. The HVM state is at the boundary of the QAH phase, whose carriers are intrinsically 100\% valley polarized.
With the in-plane magnetic anisotropy, no special QAH states and prominent valley polarization are observed. However, for both out-of-plane and in-plane magnetic anisotropy, sign-reversible  Berry curvature can be observed with increasing $U$.  It is  found that these phenomenons are related with the change of $d_{xy}$/$d_{x^2-y^2}$ and $d_{z^2}$ orbital distributions and different magnetocrystalline directions.
It is also  found that the magnetic anisotropy energy (MAE) and Curie temperature strongly depend on the $U$.  With PMA,  taking typical $U=$2.5 eV,
the electron valley polarization can be observed with valley splitting of 109 meV, which can be switched by reversing the
magnetization direction. The analysis and results  can be readily extended to other nine members of monolayer FeXY (X/Y=F, Cl, Br and I) due to sharing
the same Fe-dominated  low-energy states and electronic correlations with FeClF monolayer. Our works emphasize the importance of electronic correlation to determine the
electronic state of some materials, and the electronic correlation can induce exceptional phase transition.
\end{abstract}
\keywords{Electronic correlation, Valleytronics, Magnetic anisotropy~~~~~~~~~~~~~~~~~~~~~~~~~~~~~~~~~~~Email:sandongyuwang@163.com}

\maketitle

\section{Introduction}
In recent years, the valley as a new degree of freedom of electrons for two-dimensional
(2D) graphene-related materials has
attracted intensive attention\cite{q1,q2,q3,q4,q5,q6,q9}.
The local energy extremes in
the conduction band or valence band are referred
to as valleys, and  they provide a new effective degree of
freedom, in addition to conventional charge and spin.
 With the broken  centrosymmetry, the two inequivalent sublattices
give rise to a degenerate but inequivalent pair of valleys, which  are
well separated in the 2D hexagonal Brillouin zone\cite{q1}.
To achieve valley application,  the external conditions, such as  the optical pumping, magnetic field, magnetic
substrates and  magnetic doping\cite{q9-1,q9-2,q9-3,q9-4},  have been used to trigger  polarization.
The FV  materials may be the best choice for valleytronics due to  spontaneous valley polarization\cite{q10}, and many FV  materials have been predicted by the first-principle calculations\cite{q11,q12,q13,q13-1,q14,q15,q16,q17}.

The FV  materials possess magnetism, and generally contain transition metal
elements with localized $d$ electrons, where the electronic correlations may have important effects on  the magnetic, topological, and valley
properties of these materials. Recently, the some of   monolayer FeXY (X/Y=F, Cl, Br and I) family, such as  $\mathrm{FeX_2}$ (X=Cl, Br and I) and FeClBr\cite{v1,v2,v3,v4},  are predicted to be FV  materials. Some differences can be found for their valley properties\cite{v1,v2,v4}, which  is because different exchange-correlation functional is adopted.
 For example, for $\mathrm{FeCl_2}$, the valley polarization appears at the valence bands by using GGA\cite{v1,v4}, but that exists at the conduction bands with HSE06 and GGA+$U$ ($>$1.9 eV)\cite{v2,v4}.
  Within GGA, it is found that the valley polarization can be observed at the valence bands for  $\mathrm{FeX_2}$ (X=Cl, Br and I) and FeClBr\cite{v1,v3,v4}. These mean that the electronic correlations have important effects on physical properties of monolayer FeXY family.
 In fact, the different correlation strengths can make  $\mathrm{FeCl_2}$ monolayer  into  different ground states, like HVM and QAH states,  with the assumption of PMA\cite{v4}. However, the correlation strength of a given material is fixed. So, it is very necessary for calculating electronic properties of FeXY  family  to  consider different strength of electronic correlation.

\begin{figure*}
  \includegraphics[width=14cm]{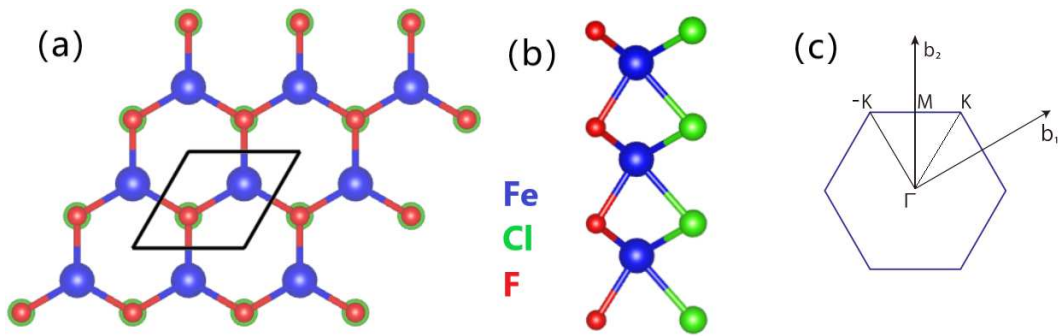}
  \caption{(Color online)The (a) top view and (b) side view of  crystal structure of Janus  monolayer  FeClF, and the rhombus primitive cell is marked by the black frame. (c) The Brillouin zone with high-symmetry points labeled.}\label{st}
\end{figure*}
\begin{figure}
  \includegraphics[width=7cm]{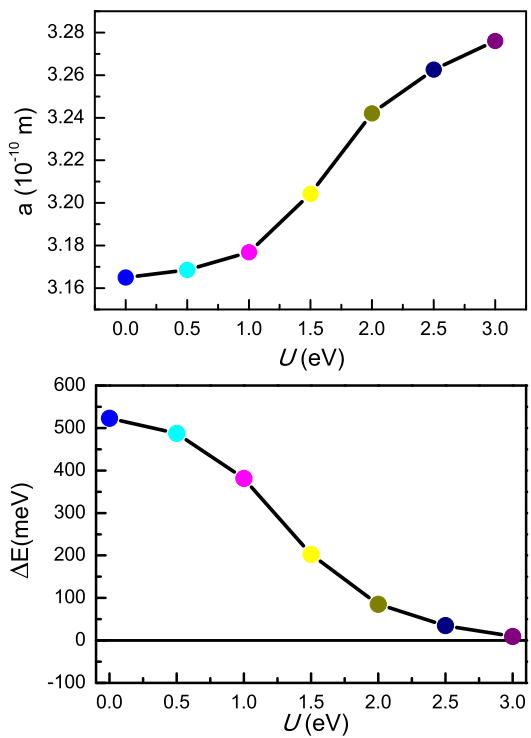}
  \caption{(Color online)For Janus monolayer FeClF, the lattice constants and energy differences (rectangle supercell) between  FM and AFM ordering  as a function of $U$.}\label{ae}
\end{figure}

In light of these  factors about monolayer FeXY family mentioned above, we  take Janus  monolayer FeClF as a concrete example to investigate  the correlation effects on its electronic structures by  GGA+$U$ approach. Unlike previous studies\cite{v4} (The lattice constants are optimized with GGA, and the electronic structures are studied by GGA+$U$.), the lattice constants are optimized with varied $U$, and the corresponding electronic structures and magnetic properties are investigated. It is found
that different  correlation strengths ($U$) along with different magnetic anisotropy (out-of-plane and  in-plane) can drive the system into
different electronic states. For PMA,  the increasing $U$ can induce the FV to HVM  to QAH to HVM to FV transitions.
 For in-plane situation, there are no special QAH states and prominent valley polarization.
These can be explained by considering  the different Fe-$d$ orbital contributions, when including spin-orbital coupling (SOC).
However, calculated results show   sign-reversible  Berry curvature with increasing $U$ for both out-of-plane and in-plane magnetic anisotropy.
 Finally, it is proved that correlation strengths have very important effects on Curie temperature $T_C$ of FeClF. The $T_C$ (311 K) with $U=$1.5 eV is
 five time that (63 K) with $U=$2.5 eV.  Our works highlight the role of
correlation effects in the 2D FeXY family materials.

The rest of the paper is organized as follows. In the next
section, we shall give our computational details and methods.
 In  the next few sections,  we shall present structure and stability, electronic structure and valley properties of Janus monolayer FeClF. Finally, we shall give our discussion and conclusion.

\section{Computational detail}
The spin-polarized  first-principles calculations  are performed   employing the projected
augmented wave (PAW) method  within density functional theory (DFT)\cite{1},  as implemented in VASP code\cite{pv1,pv2,pv3}.
 The exchange-correlation effect is treated
by the GGA of Perdew-Burke-Ernzerhof (PBE-GGA)\cite{pbe}.
The energy cut-off of 500 eV and total energy  convergence criterion of  $10^{-8}$ eV  are used in the static calculations.
The force
convergence criteria  is set to be  less than 0.0001 $\mathrm{eV.{\AA}^{-1}}$   on each atom.
The on-site Coulomb correlation of Fe atoms is considered within
the GGA+$U$ scheme by the
rotationally invariant approach proposed by Dudarev et al, in which only the effective
$U$ ($U_{eff}$) based on the difference between the on-site Coulomb interaction
parameter  and exchange parameters  is meaningful.  The SOC effect is explicitly included in the calculations to investigate MAE and electronic structures of FeClF monolayer.  A vacuum space of more than 18 $\mathrm{{\AA}}$ is used to avoid the interactions
between the neighboring slabs.
The k-mesh
of 24$\times$24$\times$1 is used to sample the Brillouin zone for calculating electronic structures and elastic properties, and 12$\times$24$\times$1 Monkhorst-Pack k-point mesh for the energies of ferromagnetic (FM) and antiferromagnetic (AFM) states with rectangle supercell, as shown in FIG.1 of electronic supplementary information (ESI).

The elastic stiffness tensor  $C_{ij}$   are calculated by using strain-stress relationship (SSR) method,
where 2D elastic coefficients $C^{2D}_{ij}$
have been renormalized by   $C^{2D}_{ij}$=$L_z$$C^{3D}_{ij}$ with the $L_z$ being  the length of unit cell along z direction.
The phonon dispersion spectrum is calculated with the 5$\times$5$\times$1 supercell by using finite displacement method,   as implemented in the  Phonopy code\cite{pv5}.  The 40$\times$40 supercell and  $10^7$ loops are used  to perform  the
Monte Carlo (MC) simulations, as implemented in Mcsolver code\cite{mc}.
 The mostly
localized Wannier functions including the $d$-orbitals of Fe atom and the  $p$-orbitals of Cl and F atoms are constructed on a k-mesh of  24$\times$24$\times$1 by the Wannier90 package\cite{w1}. The  edge states are calculated  with the software package
Wanniertools by  the renormalized effective tight binding
Hamiltonian\cite{w2}.  The Berry curvatures of FeClF monolayer
are calculated directly from the calculated
wave functions  based on Fukui's
method\cite{bm},  as implemented in the VASPBERRY code.

\begin{figure}
  \includegraphics[width=7cm]{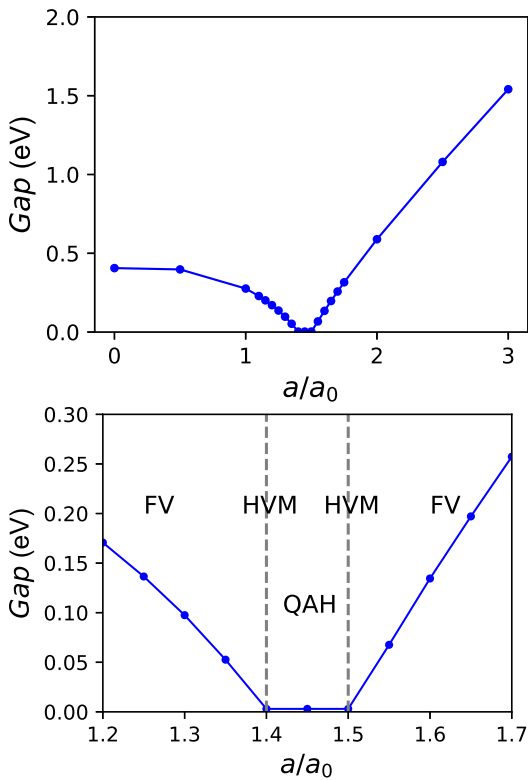}
  \caption{(Color online)For out-of-plane  magnetic anisotropy, the energy band gap  of Janus  monolayer  FeClF as a function of   $U$ (0-3 eV). The bottom plane is the enlarged view of the top plane near the $U$=1.45 eV, and   the phase diagram  with different $U$ is shown.}\label{gap}
\end{figure}

\begin{figure*}
  \includegraphics[width=14cm]{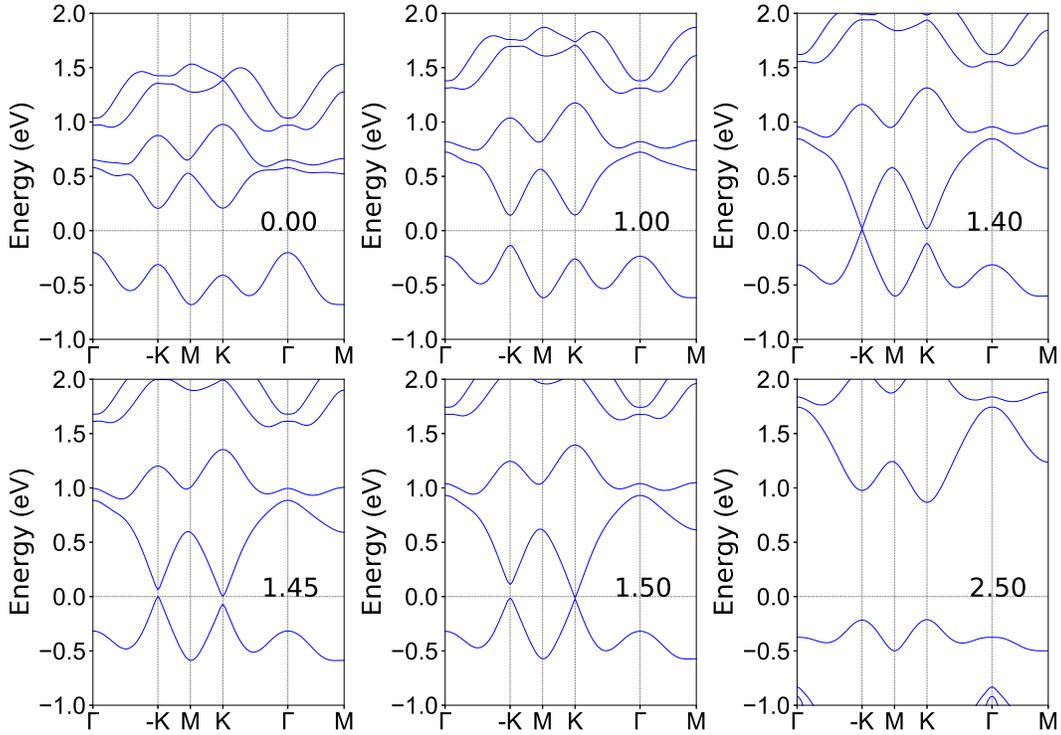}
\caption{(Color online)For out-of-plane magnetic anisotropy, the  energy band structures  of Janus  monolayer  FeClF   with $U$ being 0.00 eV, 1.00 eV, 1.40 eV, 1.45 eV, 1.50 eV and 2.50 eV.}\label{band}
\end{figure*}

\begin{figure*}
  \includegraphics[width=13cm]{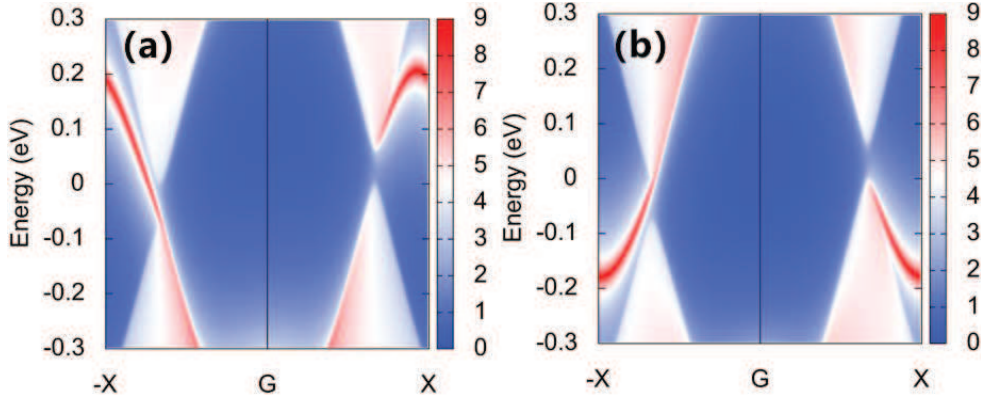}
\caption{(Color online)For out-of-plane  magnetic anisotropy,  topological
edge states of Janus monolayer FeClF calculated  along the (100) direction with $U$ being 1.45 eV, including  left (a) and right (b)  edges.}\label{s-1}
\end{figure*}

\section{Structure and stability}
 The crystal structures of the Janus FeClF
monolayer are plotted  in \autoref{st}, along with Brillouin zone with high-symmetry points.
It is clearly seen that the FeClF monolayer  consists of Cl-Fe-F sandwich layer, which can be built
by  replacing one of two  Cl   layers with F  atoms in  $\mathrm{FeCl_2}$  monolayer.
In experiment,  Janus monolayer MoSSe can be achieved from $\mathrm{MoS_2}$ by replacing one of two  S   layers with Se  atoms\cite{e1,e2}.
Due to broken vertical mirror symmetry,  the space group of  FeClF monolayer is $P3m1$ (No.156), which is  lower than $P\bar{6}m2$ of  $\mathrm{FeCl_2}$ monolayer (No.187). The FeClF ($\mathrm{FeCl_2}$) has the same symmetry with MoSSe ($\mathrm{MoS_2}$). The lattice constants $a$ is optimized with different $U$ (0-3 eV), which is plotted in \autoref{ae}. It is found that the $a$ increases with increasing $U$, and ranges from 3.165 $\mathrm{{\AA}}$ to 3.276 $\mathrm{{\AA}}$. To determine magnetic ground state, the energy differences between AFM and FM ordering as a function of $U$ are also plotted in \autoref{ae}. In considered $U$ range, the FM order is the most stable magnetic state.  It is found that the FM interaction decreases with increasing $U$, which can produce important effects on Curie temperature of FeClF monolayer.

To demonstrate the stability of Janus FeClF monolayer, the phonon spectra and elastic constants are calculated by using GGA method.
The phonon band dispersions of FeClF monolayer calculated along the high-symmetry
directions of the Brillouin zone without imaginary frequency modes are plotted  in FIG.1 of ESI, suggesting its dynamical stability.
To check
the mechanical stability of monolayer  FeClF,  the elastic
properties  are investigated.
Due to $P3m1$ space group,  using Voigt notation, the elastic tensor can be reduced into:
\begin{equation}\label{pe1-4}
   C=\left(
    \begin{array}{ccc}
      C_{11} & C_{12} & 0 \\
     C_{12} & C_{11} &0 \\
      0 & 0 & (C_{11}-C_{12})/2 \\
    \end{array}
  \right)
\end{equation}
The independent $C_{11}$ and $C_{12}$  of FeClF monolayer are 68.88 $\mathrm{Nm^{-1}}$ and 22.21 $\mathrm{Nm^{-1}}$.  These elastic constants satisfy the  Born  criteria of mechanical stability\cite{ela}: $C_{11}$$>$0 and $C_{11}-C_{12}$$>$0,   confirming its  mechanical stability.
The  Young¡¯s moduli $C^{2D}$, shear modulus $G^{2D}$ and Poisson's ratios $\nu^{2D}$ of FeClF monolayer  are mechanically isotropic due to hexagonal symmetry, and the corresponding values are 61.72 $\mathrm{Nm^{-1}}$, 23.34 $\mathrm{Nm^{-1}}$ and 0.32.

\section{Out-of-plane magnetic anisotropy}
Firstly, we consider that the magnetocrystalline direction  of
 FeClF monolayer is along the out-of-plane.  The out-of-plane FM maintains the horizontal mirror symmetry, and breaks all
possible vertical mirrors, which  allows a nonvanishing Chern
number of the 2D system\cite{n1}.  The evolutions of electronic band
structures with $U$ are investigated by GGA+SOC. The energy band gaps as a function of $U$ are plotted in \autoref{gap}. The representative energy band structures at different U
values are shown in \autoref{band}. When $U<$1.4 eV, the gap decreases with increasing $U$. However, with increasing $U$, the gap increases for $U>$1.5 eV. Between $U$=1.4 eV and 1.5 eV, very little gap can be observed, which may show nontrivial topological properties.

\begin{figure*}
  \includegraphics[width=16cm]{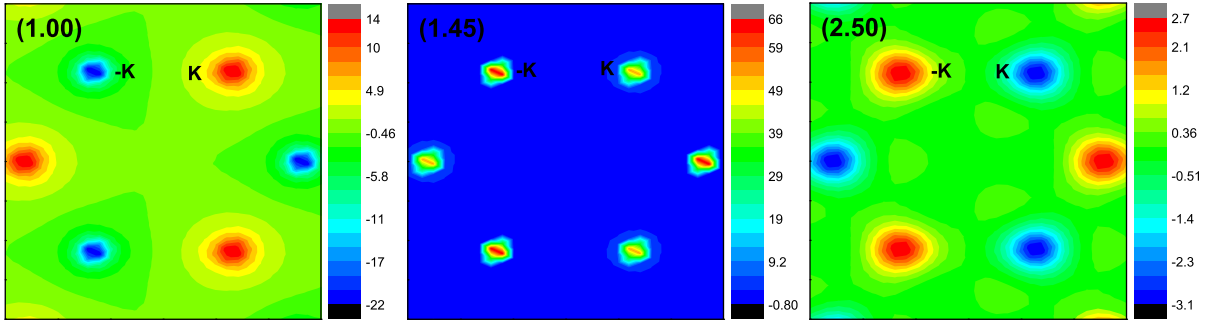}
  \caption{(Color online)For out-of-plane  magnetic anisotropy, the calculated Berry curvature distribution of Janus  monolayer FeClF in the 2D Brillouin zone with $U$ being  1.00 eV, 1.45 eV and 2.50 eV.}\label{berry}
\end{figure*}

For $U<$1.4 eV, a remarkable valley polarization can be observed in the  valence bands at the -K and K points, and the  -K point is polarized. However, for $U>$1.5 eV,  the noteworthy valley polarization occurs in the conduction bands, and the K point is polarized.
In these two regions, the monolayer FeClF is a FV  material.
Around the $U$=1.40 eV,  the band gap of -K point  gets closed, and a narrow band gap still holds at K point,  signifying the
HVM, whose  conduction electrons are intrinsically 100\% valley polarized\cite{v4}. The HVM can also be observed around the $U$=1.5 eV, but the band gap  at K point disappears, and a narrow band gap  exists at -K point.
 The gap closes,
reopens, and then closes, which suggests a possible topological phase transition.  Between the two HVM states,  a
QAH insulator phase may exist, which can be characterized by   chiral edge states. The edge states are calculated along (100) direction with $U$=1.45 eV, which is plotted in \autoref{s-1}. It is clearly seen that there does exist a chiral edge state connecting the conduction bands (-K/K valley) and
valence  bands (K/-K valley) for left/right edge.   A single gapless chiral edge band means that the Chern number is equal to one ($C$=1), which is consistent with integration over the Berry curvatures (See \autoref{berry}),  giving a nonzero Chern number ($C$=1).
It is found  that the QAH phase  coexists with a valley structure for both conduction and valence bands.
When shifting the Fermi level into conduction/valence bands,  the K/-K valley is polarized.
In a word, the FeClF monolayer  undergoes the FV, HVM, QAH, HVM and FV states with increasing $U$ (see \autoref{gap}).

\begin{figure}
   \includegraphics[width=8cm]{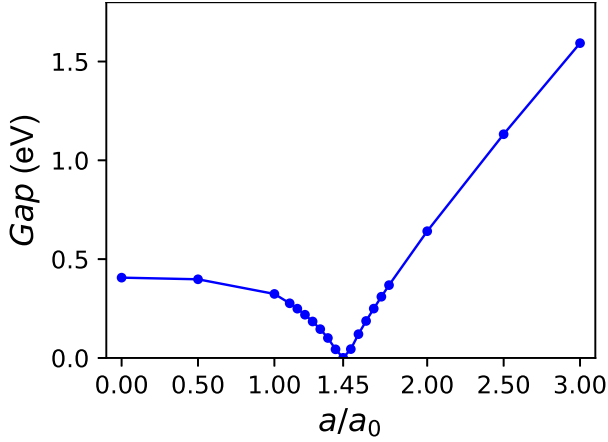}
  \caption{(Color online)For  in-plane magnetic anisotropy, the energy band gap  of Janus  monolayer  FeClF as a function of   $U$ (0-3 eV).}\label{gap-1}
\end{figure}
\begin{figure*}
  \includegraphics[width=15cm]{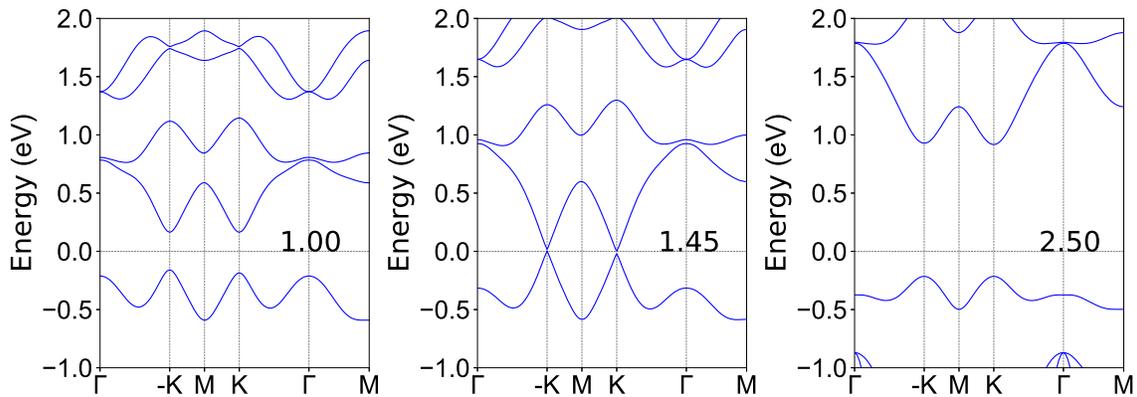}
\caption{(Color online)For  in-plane magnetic anisotropy, the  energy band structures  of Janus  monolayer  FeClF   with $U$ being  1.00 eV,  1.45 eV and 2.50 eV.}\label{band-1}
\end{figure*}

\begin{figure}
   \includegraphics[width=8cm]{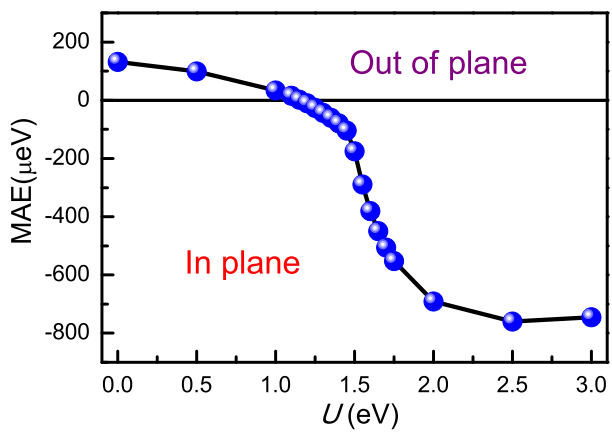}
  \caption{(Color online) The MAE of Janus  monolayer  FeClF as a function of   $U$ (0-3 eV).}\label{m}
\end{figure}
\begin{figure*}
  \includegraphics[width=15cm]{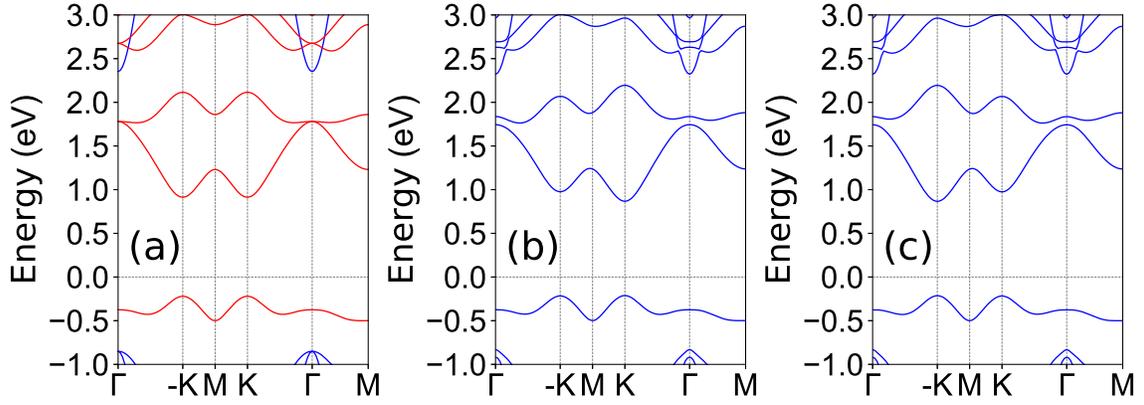}
\caption{(Color online)For out-of-plane magnetic anisotropy with $U$=2.5 eV,  the energy  band structures of  Janus  monolayer  FeClF (a) without SOC; (b) and (c) with SOC for magnetic moment of Fe along the positive and negative z direction, respectively.  In (a), the blue (red) lines represent the band structure in the spin-up (spin-down) direction.}\label{band-2}
\end{figure*}

\begin{figure}
   \includegraphics[width=6cm]{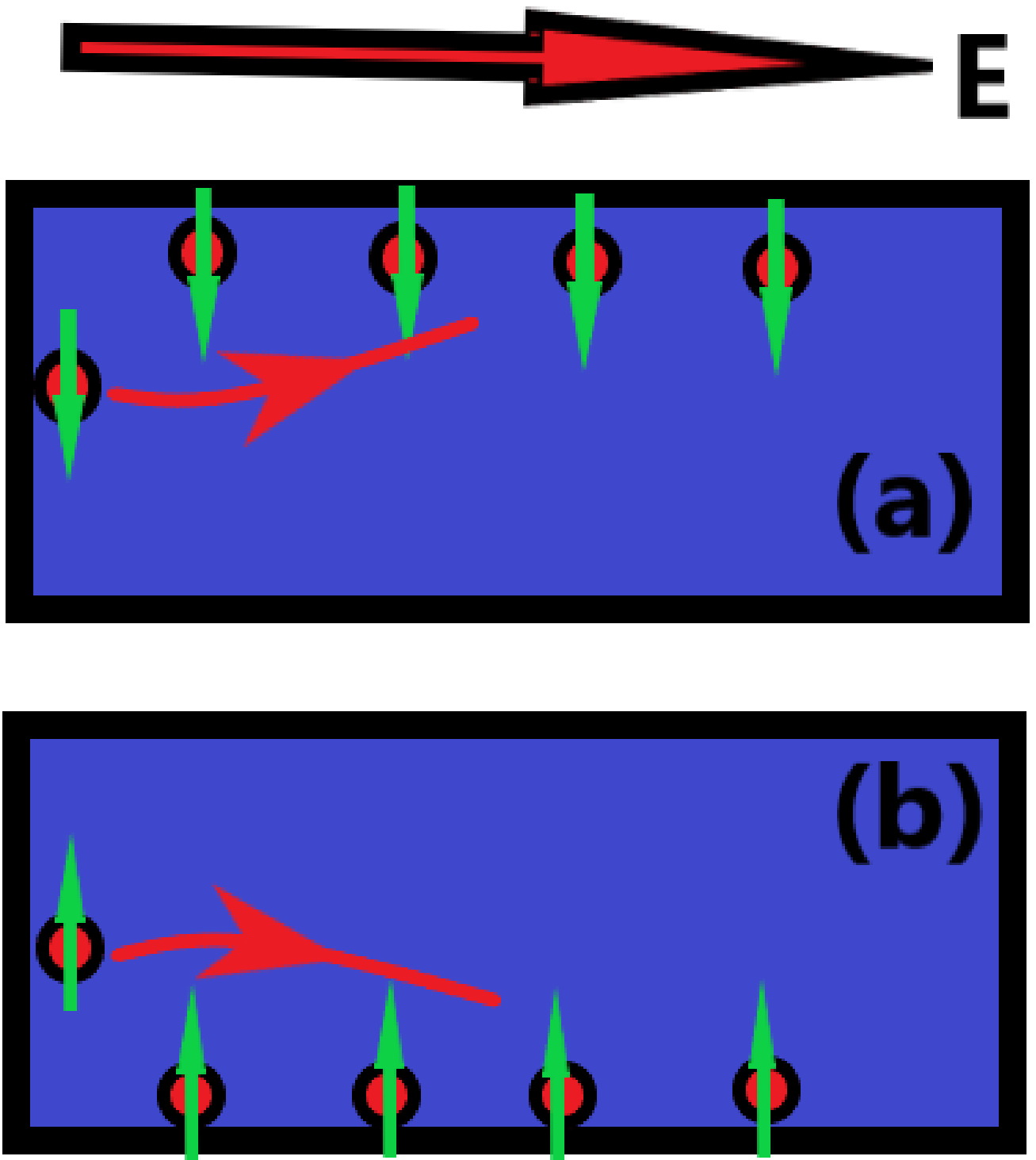}
  \caption{(Color online)(a) The schematics of anomalous valley Hall effect under electron doping. (b) is the same as (a) but with opposite magnetic moment. The green  arrows  represent spin-down or spin-up states.}\label{va}
\end{figure}

Berry curvature  is a powerful tool to investigate the valley physics.  The Berry curvature is calculated with $U$ from 0 to 3 eV, and the representative ones at different $U$
values are plotted in \autoref{berry}. For $U$=1.00 eV, the
hot spots in the Berry curvature  are around two valleys, which have opposite
signs and different magnitudes.   With increasing $U$,
the FV state changes into the
QAH state by a topological phase transition, connected by the HVM state. Within the range of $U$=1.4-1.5 eV, the sign of Berry curvature  at -K valley
flips (For example $U=$1.45 eV in \autoref{berry}).  When $U$ spans the 1.5 eV, the QAH state changes into FV state with middle HVM state, resulting
in the sign change of Berry curvature at K valley (For example $U=$2.50 eV in \autoref{berry}). These mean that sign-reversible  Berry curvature can be observed with increasing $U$.

For $U<$1.4 eV, the valley polarization is in the valence bands, but  the valley polarization is in the conduction bands  for $U>$1.5 eV.
For $U$ between 1.4 eV and 1.5 eV, the valley polarization can exit in both valence and conduction bands.
To explain the peculiar effect of electronic correlation and SOC on the band
structure, the Fe-$d$-orbital characters of energy bands for $U=$1.00 eV, 1.45 eV and 2.50 eV are plotted in FIG.2 of ESI.  For all considered $U$,  the -K and K valleys in both valence and conduction bands are dominated by $d_{z^2}$ or $d_{x^2-y^2}$/$d_{xy}$ orbitals. For $U<$1.4 eV, the -K and K valleys in valence bands are dominated by $d_{x^2-y^2}$ and $d_{xy}$ orbitals, and those in the conduction bands  are
mainly from the $d_{z^2}$ orbitals.  For $U>$1.5 eV, the opposite situation can be observed. When $U$ is between  1.4 eV and 1.5 eV, the -K valley in the conduction bands and K valley in the valence band are dominated by $d_{x^2-y^2}$ and $d_{xy}$ orbitals, and the -K valley in the valence  bands and K valley in the conduction band are dominated by $d_{z^2}$ orbitals.

The valley polarization induced by SOC  is due to  the intra-atomic interaction:
\begin{equation}\label{m1}
\hat{H}_{SOC}=\lambda\hat{L}\cdot\hat{S}=\hat{H}^0_{SOC}+\hat{H}^1_{SOC}
\end{equation}
in which $\lambda$ is the coupling strength, and  $\hat{L}$ and $\hat{S}$ are the orbital angular moment and spin angular moment, respectively.
The magnetic exchange interaction results in that  $\hat{H}^1_{SOC}$ (the  interaction of opposite spin states) can be  ignored.
So, the $\hat{H}^0_{SOC}$ (the interaction
between the same spin states)  dominates the  $\lambda\hat{L}\cdot\hat{S}$.
The  $\hat{H}^0_{SOC}$ can be written as\cite{q10,v2,v3}:
\begin{equation}\label{e1}
\hat{H}^0_{SOC}=\lambda\hat{S}_{z^`}(\hat{L}_z cos\theta+\frac{1}{2}\hat{L}_{+}e^{-i\phi}sin\theta+\frac{1}{2}\hat{L}_{-}e^{+i\phi}sin\theta)
\end{equation}
 where $\theta$ and $\phi$ are the polar angles of spin orientation.  For out-of-plane magnetization ($\theta=0$$^{\circ}$),  $\hat{H}^0_{SOC}$  can be reduced to:
\begin{equation}\label{m1}
\hat{H}^0_{SOC}=\alpha \hat{L}_z
\end{equation}
At -K and K valleys, the group symmetry is $C_{3h}$.  Hence, the orbital
basis for -K and K valleys can be expressed as\cite{q10,v2,v3}:
\begin{equation}\label{m2}
   \begin{array}{c}
|\phi^\tau>=\sqrt{\frac{1}{2}}(|d_{x^2-y^2}>+i\tau|d_{xy}>)\\
or\\
|\phi^\tau>=|d_{z^2}>
  \end{array}
\end{equation}
where the subscript  $\tau$ represent  valley index ($\tau=\pm1$).  The resulting energy  at K
and -K valleys can be expressed as:
\begin{equation}\label{m3}
E^\tau=<\phi^\tau|\hat{H}^0_{SOC}|\phi^\tau>
\end{equation}
If the -K and K valleys are dominated by $d_{x^2-y^2}$ and $d_{xy}$ orbitals, the valley  energy difference $|\Delta E|$
 at -K and K points (valley splitting) can be expressed as:
\begin{equation}\label{m4}
|\Delta E|=E^{K}-E^{-K}=4\alpha
\end{equation}
If the -K and K valleys are
mainly from the $d_{z^2}$ orbitals, the valley splitting $|\Delta E|$
 is given by:
\begin{equation}\label{m4}
|\Delta E|=E^{K}-E^{-K}=0
\end{equation}
So, these different orbital components of -K and K valleys with varied $U$ determine the valley polarization distribution.

\section{in-plane magnetic anisotropy}
Next, we consider that the magnetocrystalline direction  of
 FeClF monolayer is along the in-plane.  The  electronic band
structures with different  $U$ are calculated by GGA+SOC. The energy band gaps vs $U$ are shown in \autoref{gap-1}, and the representative energy band structures  are plotted in \autoref{band-1}. It is found that the energy band gap firstly decreases, and then increases, when the $U$ increases.
The critical $U$ value is about 1.45 eV. Compared to out-of-plane magnetocrystalline direction,  no special  intermediate region exists.
\autoref{band-1} shows no observable valley polarization  in both the valence and conduction bands, which  can be explained by $\Delta E=4\alpha cos\theta$\cite{v3}. When the magnetocrystalline direction  of
 FeClF monolayer is along in-plane direction ($\theta$=90$^{\circ}$), the valley splitting will vanish ($\Delta E$=0).
The Fe-$d$-orbital characters of energy bands for $U=$1.00 eV, 1.45 eV and 2.50 eV are plotted in FIG.3 of ESI. It is found that the distributions of Fe-$d$-orbital characters are akin to the cases of out-of-plane. According to FIG.4 of ESI, the sign-reversible Berry curvature can be observed, when the $U$ value strides over the
critical point (about $U=$1.45 eV).  Calculated results show that the
hot spots in the Berry curvature  are around two valleys with opposite
signs and almost the same magnitudes.  To explore the topological properties of FeClF monolayer at different $U$, we calculate the dispersion of the edge state. The edge states  with  representative $U$ are plotted in FIG.5 of ESI (only left edge), which show that no chiral edge states traverse the bulk band
gap. This mean that, for in-plane  magnetocrystalline direction, no QAH states appear. So, the magnetocrystalline direction is very important to investigate electronic properties of FeClF monolayer.

\section{Magnetic anisotropy energy and anomalous valley Hall effect}
The magnetocrystalline direction  of
 FeClF monolayer can be regulated by external magnetic field. However, we use MAE  to determine  intrinsic magnetic anisotropy of FeClF monolayer at different $U$ value. Within GGA+SOC+$U$, the MAE can be calculated by $E_{MAE}$ = $E_{(100)}$-$E_{(001)}$, which is plotted
 in \autoref{m}.
The positive value means that the easy axis  is perpendicular to the plane, while the negative value suggests the in-plane direction.
For $U<$1.15 eV, the FeClF monolayer possesses out-of-plane magnetic anisotropy. For $U>$1.15 eV, the direction of the easy axis of FeClF monolayer  is in plane. We investigate the valley  properties of FeClF monolayer with  representative $U$ (2.5 eV)\cite{fe,fe1}, and its magnetocrystalline direction can be switched to out-of-plane by external magnetic field.  The spin-polarized band structures of monolayer  FeClF without and with SOC are shown in \autoref{band-2}.
According to \autoref{band-2} (a),  a distinct
spin splitting can be observed in the band structures because of the exchange
interaction, and  the FeClF monolayer is  a direct band
gap semiconductor with VBM and CBM  provided by the same spin-down. The valleys of -K and K are degenerate in energy
for both conduction and valence bands. \autoref{band-2} (b) shows that the SOC effect can induce valley polarization in the conduction bands.
The valley splitting $|\Delta E|$ is 109  meV,  and   the energy of K valley
is lower than one of -K valley. Moreover, the valley polarization can  be
switched by reversing the magnetization direction with magnetic moment of Fe along the negative z direction, which
is confirmed by  \autoref{band-2} (c). Because  the low-energy bands at -K and K belong
to the same  spin minority channel, the spin polarization of the carriers is simultaneously switched.

It is also known that  Berry curvature $\Omega(k)$  is associated with anomalous velocity $\upsilon$  of Bloch electrons  under an in-plane longitudinal electric field $E$: $\upsilon\sim E\times\Omega(k)$\cite{q9-3}. According to \autoref{berry}, Berry curvature $\Omega(k)$ is characterized with unequal values and opposite signs for the -K and K valleys. With the reversed opposite magnetic moment, the valley polarization will also be reversed.  The corresponding values of the Berry curvatures at the K and -K valleys will be exchanged, but  their sign  remains  unchanged.
Under such condition, the anomalous valley Hall effect can be observed in monolayer FeClF. When shifting the Fermi level between the K and -K valleys, the spin-down electrons at K valley will accumulate on the one  side  of the sample[\autoref{va} (a)].
By reversing the magnetization direction, the spin-up electrons at -K valley will gain opposite transverse velocities, moving towards another side [\autoref{va} (b)] of the sample due to its opposite Berry curvature.

\begin{figure*}
  \includegraphics[width=12cm]{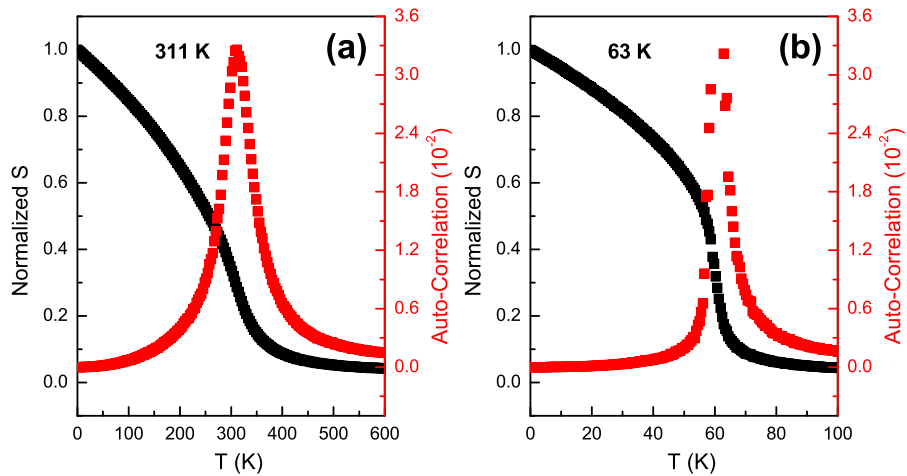}
  \caption{(Color online)For Janus monolayer FeClF, the normalized magnetic moment (S) and auto-correlation  as a function of temperature with $U=$1.5 eV (Left) and 2.5 eV (Right).}\label{ae-1}
\end{figure*}

\section{Curie temperature}
We also estimate the Curie temperature $T_C$ of monolayer FeClF at  representative U values by MC simulations with the
Wolf   algorithm based on the Heisenberg model. The effective classical spin
model  can be written as:
  \begin{equation}\label{pe0-1-1}
H=-J\sum_{i,j}S_i\cdot S_j-A\sum_i(S_i^z)^2
 \end{equation}
where  $S_i$/$S_j$, $S_i^z$, $J$ and  $A$  are   the
spin vectors of each Fe atom,  the spin component parallel to the z direction,  the nearest neighbor exchange parameter and   MAE, respectively. To extract $J$, we compare energies  of the FM  ($E_{FM}$) and AFM ($E_{AFM}$)
 configurations of monolayer FeClF with rectangle supercell. The spin vector is normalized ($|S|$ = 1), and the corresponding energies are given by:
  \begin{equation}\label{pe0-1-2}
E_{FM}=E_0-6J-2A
 \end{equation}
  \begin{equation}\label{pe0-1-3}
E_{AFM}=E_0+2J-2A
 \end{equation}
 where $E_0$ is the energy without magnetic coupling. According to these equations, the  $J$ can be attained as:
  \begin{equation}\label{pe0-1-3}
J=\frac{E_{AFM}-E_{FM}}{8}
 \end{equation}
The calculated normalized $J$ at $U=$ 1.5 eV and 2.5 eV are  25.30 meV and 4.40 meV. The   normalized magnetic moment and auto-correlation of monolayer FeClF  vs temperature are shown in \autoref{ae-1}, and the predicted $T_C$ is about 311 K and 63 K. These mean that electron correlation can produce important effects on  Curie temperature, which is also can be understood by energy differences  between  FM and AFM ordering vs $U$ (see \autoref{ae}).  The Curie temperature of FeClBr is predicted to be very high (651 K)\cite{v3}, which is due to adopt GGA method ($U$=0 eV).

\section{Discussion and Conclusion}
The importance of electron correlations on the electronic  properties of monolayer FeClF has been  demonstrated by the first-principles calculations.
The different correlation strength  can  give rise to different electronic states, like FV, HVM and QAH states.  Although the  monolayer FeClF as a concrete example is investigated,  the analysis and results in the work  can be readily extended
to other members of monolayer FeXY (X/Y=F, Cl, Br and I), including ten monolayers. This is because the low-energy states and the
electronic correlations are dominated by Fe atoms, and the rich correlation-driven physics
should be common among the FeXY family.  Similar phase diagram of monolayer $\mathrm{FeCl_2}$ with varied $U$ has been investigated\cite{v4}, where the out-of-plane magnetocrystalline direction is always assumed.  The spontaneous valley
polarization of monolayer  FeClBr  has also been studied by using GGA, and  the MAE is 14 $\mathrm{\mu eV}$/Fe with out-of-plane direction\cite{v3}, which is lower than one (132 $\mathrm{\mu eV}$/Fe) of FeClF within GGA. The predicted Curie
temperature of FeClBr monolayer is very high (651 K)\cite{v3}, which should be reduced with increasing $U$ according to our calculated results.
The correlation-driven topological and valley states in  septuple atomic  monolayer $\mathrm{VSi_2P_4}$ have been revealed\cite{v5}, whose intermediate three layers ($\mathrm{VP_2}$) share the similar structure with FeXY family. For a given material, the correlation strength should be constant, which should
be determined from experiment. However, the rich electronic states can be realized by tuning  correlation effect, which can be achieved by  applied strain. The electronic correlation depends on the competition between kinetic and interaction energies, and the strain  can modify the bandwidth, adjust correlation effect\cite{v5}. The rich  phase diagram in septuple atomic  monolayer $\mathrm{VSi_2N_4}$  has been achieved by strain\cite{v6}, and sign-reversible valley-dependent Berry phase effects and QAH states  have  been investigated.

In summary, we have demonstrated the significance of electron correlation
 in Janus monolayer FeClF, as a representative of the
2D  FeXY material family.
It is found  that different correlation strength (varied $U$) can result in different electronic state due to  interplay between magnetic,
correlation and SOC. The multiple transitions  are observed, including  the magnetic anisotropy, valley structure, electroconductibility
and electronic topology. For PMA, there exists a QAH phase, whose boundary  corresponds to the HVM state with fully valley
polarized carriers. For FV phases, the polarization can  exist for
both electrons (conduction bands) and holes (valence bands),  which depends the correlation strengths.
The polarization at -K and K valleys can be switched by reversing
the magnetization. The increasing $U$ can induce  sign-reversible  Berry curvature for both out-of-plane and in-plane magnetic anisotropy.
The magnetic interaction related with Curie temperature can be distinctly influenced by correlation strength. Our works provide a comprehensive understanding  of
the correlation effects in the 2D FeXY family, which  can be spread to other 2D FV  materials.

\begin{acknowledgments}
This work is supported by Natural Science Basis Research Plan in Shaanxi Province of China  (2021JM-456). We are grateful to the Advanced Analysis and Computation Center of China University of Mining and Technology (CUMT) for the award of CPU hours and WIEN2k/VASP software to accomplish this work.
\end{acknowledgments}

\end{document}